\documentclass[10pt,conference,final]{IEEEtran}
\IEEEoverridecommandlockouts

\usepackage[utf8]{inputenc}
\usepackage{cite}
\usepackage{amsmath,amssymb,amsfonts}
\usepackage{algorithmic}
\usepackage{graphicx}
\usepackage{textcomp}
\usepackage{xcolor}

\usepackage{color}
\definecolor{pblue}{rgb}{0.13,0.13,1}
\definecolor{pgreen}{rgb}{0,0.5,0}
\definecolor{pred}{rgb}{0.9,0,0}
\definecolor{pgrey}{rgb}{0.46,0.45,0.48}

\usepackage[scaled]{beramono}
\usepackage{tikz}
\usepackage{xspace}
\usepackage{listings}
\makeatletter
\newenvironment{btHighlight}[1][]
{\begingroup\tikzset{bt@Highlight@par/.style={#1}}\begin{lrbox}{\@tempboxa}}
{\end{lrbox}\bt@HL@box[bt@Highlight@par]{\@tempboxa}\endgroup}
\newcommand\btHL[1][]{%
  \begin{btHighlight}[#1]\bgroup\aftergroup\bt@HL@endenv%
}
\def\bt@HL@endenv{%
  \end{btHighlight}%
  \egroup
}
\newcommand{\bt@HL@box}[2][]{%
  \tikz[#1]{%
    \pgfpathrectangle{\pgfpoint{1pt}{0pt}}{\pgfpoint{\wd #2}{\ht #2}}%
    \pgfusepath{use as bounding box}%
    \node[anchor=base west, fill=orange!30,outer sep=0pt,inner xsep=1pt, inner ysep=0pt, rounded corners=3pt, minimum height=\ht\strutbox+1pt,#1]{\raisebox{1pt}{\strut}\strut\usebox{#2}};
  }%
}
\lstdefinestyle{Java}{
    language={Java}, basicstyle=\ttfamily\footnotesize, 
    moredelim=**[is][{\btHL[fill=red!17,thin]}]{`}{`},
    moredelim=**[is][{\btHL[fill=green!17,thin]}]{@}{@},
    moredelim=**[is][{\btHL[fill=yellow!17,thin]}]{~}{~},
}

\usepackage{algorithm}
\usepackage{dsfont}
\usepackage{lineno}
\usepackage[T1]{fontenc}
\usepackage[flushleft]{threeparttable}
\usepackage{multicol}
\usepackage{xspace}
\usepackage{amssymb}
\usepackage{amsthm}
\usepackage{lscape}
\usepackage{url}
\usepackage{multirow}
\usepackage{array}
\usepackage{paralist}
\usepackage{makecell}
\usepackage{graphicx}
\usepackage{tabu}
\usepackage[framemethod=tikz]{mdframed}
\newtheorem{defn}{Definition}

\mdfdefinestyle{mpdframe}{
    frametitlebackgroundcolor   =black!15,
    frametitlerule              =true,
    roundcorner                 =5pt,
    middlelinewidth             =0.8pt,
    innermargin                 =0.2cm,
    outermargin                 =0.2cm,
    innerleftmargin             =0.2cm,
    innerrightmargin            =0.2cm,
    innertopmargin              =0.2cm,
    innerbottommargin           =0.2cm
}

\usepackage[colorinlistoftodos,prependcaption]{todonotes}
\newboolean{showcomments}
\setboolean{showcomments}{true}
\ifthenelse{\boolean{showcomments}}
 { \newcommand{\mynote}[2]{
      \fbox{\bfseries\sffamily\scriptsize#1}
        {\small$\blacktriangleright$\textsf{\textcolor{red}{{\em #2}\bf }}$\blacktriangleleft$}}}
        { \newcommand{\mynote}[2]{}}
\usepackage{tabularx}
\newcolumntype{C}{>{\centering\arraybackslash}X}
\newcolumntype{R}{>{\raggedleft\arraybackslash}X}

\usepackage{booktabs}

\definecolor{mymauve}{rgb}{0.58,0,0.82}
\definecolor{mygrey}{rgb}{0.43, 0.5, 0.5}

\usepackage{pbox}

\input{utils/util.tex}

\usepackage[hidelinks, bookmarks=false]{hyperref}

\def\BibTeX{{\rm B\kern-.05em{\sc i\kern-.025em b}\kern-.08em
    T\kern-.1667em\lower.7ex\hbox{E}\kern-.125emX}}

\title{
The Strengths and Behavioral Quirks of\\ Java Bytecode Decompilers
}

\author{
\IEEEauthorblockN{Nicolas Harrand, C\'esar Soto-Valero, Martin Monperrus, and Benoit Baudry}
\textit{KTH Royal Institute of Technology, Stockholm, Sweden}\\ Email: \{harrand, cesarsv, baudry\}@kth.se, martin.monperrus@csc.kth.se

}

\begin{document}

\maketitle

\begin{abstract}
During compilation from Java source code to bytecode, some information is irreversibly lost. 
In other words, compilation and decompilation of Java code is not symmetric.
Consequently, the decompilation process, which aims at producing source code from bytecode, must establish some strategies to reconstruct the information that has been lost. 
Modern Java decompilers tend to use distinct strategies to achieve proper decompilation.
In this work, we hypothesize that the diverse ways in which bytecode can be decompiled has a direct impact on the quality of the source code produced by decompilers. 

We study the effectiveness of eight Java decompilers with respect to three quality indicators: syntactic correctness, syntactic distortion and \semi.
This study relies on a benchmark set of 14 real-world open-source software projects to be decompiled (2041 classes in total). 

Our results show that no single modern decompiler is able to correctly handle the variety of bytecode structures coming from real-world programs. Even the highest ranking decompiler in this study  produces syntactically correct output for 84\% of classes of our dataset and semantically equivalent code output for 78\% of classes.

\end{abstract}


\let\url\nolinkurl

\section{Introduction}\label{sec:introduction}

In the Java programming language, source code is compiled into an intermediate stack-based representation known as bytecode, which is  interpreted by the Java Virtual Machine (JVM). In the process of translating source code to bytecode, the compiler performs various analyses. Even if most optimizations are typically performed at runtime by the just-in-time (JIT) compiler, several pieces of information residing in the original source code are already  not present in the bytecode anymore due to compiler optimization~\cite{Lindholm2014}. For example the structure of loops is altered and local variable names may be modified~\cite{Jaffe2018}.

Decompilation is the inverse process, it consists in transforming the  bytecode instructions into source code~\cite{Nolan2004}. 
Decompilation can be done with several goals in mind.
First, it can be used to help developers understand the code of the libraries they use. This is why Java IDEs such as IntelliJ and Eclipse include built-in decompilers to help developers analyze the third-party classes for which the source code is not available. In this case, the readability of the decompiled code is paramount.
Second, decompilation may be a preliminary step before another compilation pass, for example with a different compiler. In this case, the main goal is that the decompiled code is syntactically and grammatically correct and can be recompiled.
Some other applications of decompilation with slightly different criteria include clone detection~\cite{Ragkhitwetsagul2017}, malware analysis ~\cite{Yakdan2016, Durfina2013} and software archaeology~\cite{Robles2005}.

Overall, the ideal decompiler is one that transforms all inputs into source code that faithfully reflects the original code: the decompiled code 1) can be recompiled with a Java compiler and 2) behaves the same as the original program. 
However, previous studies having compared Java decompilers \cite{Hamilton2009,Kostelansky2017} found that this ideal Java decompiler does not exist, because of the irreversible data loss that happens during compilation. Yet, the experimental scale of this previous work is rather small to fully understand the state of decompilation for Java. There is a fundamental reason for this: this previous work relies on manual analysis to assess the semantic correctness of the decompiled code.

In this paper, we solve this problem by proposing a fully automated approach to study Java decompilation, based on \textit{equivalence modulo inputs} (EMI) \cite{Le2014}. The idea is to automatically check that decompiled code behaves the same as the original code, using inputs provided by existing application test suites. In short, the decompiled code of any arbitrary class $x$ should pass all the tests that exercise $x$. To our knowledge, this is the first usage of EMI in the context of decompilation.
With that instrument, we perform a comprehensive assessment of three aspects of decompilation: the syntactic correctness of the decompiled code (the decompiled code can recompile); the semantic equivalence modulo input with the original source (the decompiled code passes all tests); the syntactic similarity to the original source (the decompiled source looks like the original). To our knowledge, this is the first deep study of those three aspects together.

Our study is based on $14$ open-source projects totaling $2041$ Java classes. We evaluate eight recent and notable decompilers on code produced by two different compilers. This study is at least one order of magnitude larger than the related work \cite{Hamilton2009,Kostelansky2017}. 
Our results are important for different people: 
1) for all users of decompilation, our paper shows significant differences between decompilers and provide well-founded empirical evidence to choose the best ones;
2) for researchers in decompilation, our results shows that the problem is not solved, and we isolate a subset of $157$ Java classes that no state-of-the-art decompiler can correctly handle:
3) for authors of decompilers, our experiments have identified bugs in their decompilers ($2$ have already been fixed, and counting) and our methodology of \semi can be embedded in the QA process of all decompilers in the world.

In summary, this paper makes the following contributions:
\begin{itemize}
    \item an adaptation of equivalence modulo inputs \cite{Le2014} in the context of decompiler validation;
    \item a fully automated pipeline to assess the syntactic and semantic quality of source code generated by Java decompilers;
    \item an empirical comparison of eight Java decompilers based on $2041$ real-world Java classes, tested by $25019$ test cases, identifying the key strengths and limitations of bytecode decompilation. 
    \item a tool and a dataset, publicly available for future research on Java decompilers.\footnote{\url{https://github.com/castor-software/decompilercmp}}
\end{itemize}

\section{Motivating Example}\label{sec:background}

In this section, we present an example drawn from the Apache \texttt{commons-codec} library. We wish to illustrate information loss during compilation of Java source code, as well as the different strategies that bytecode decompilers adopt to cope with this loss when they generate source code. \autoref{lst:commons-codec-util-original-src} shows the original source code of the utility class \texttt{org.apache.commons.codec.net.Utils}, while \autoref{lst:commons-codec-util-bytecode} shows an excerpt of the bytecode produced by the standard \javac compiler.\footnote{There are various Java compilers available, notably Oracle \javac and Eclipse \ecj, which can produce different bytecode for the same Java input.} Here, we omit the constant pool as well as the table of local variables and replace references towards these tables with comments to save space and make the bytecode more human readable. 
 
\begin{lstlisting}[style=Java, float, floatplacement=H, caption={Source code of \texttt{org.apache.commons.codec.net.Utils}.}, label={lst:commons-codec-util-original-src}] 
class Utils {
    private static final int RADIX = 16;
    static int digit16(final byte b) throws DecoderException {
        final int i = Character.digit((char) b, RADIX);
        if (i == -1) {
            throw new DecoderException("Invalid URL encoding: not a valid digit (radix " + RADIX + "): " + b);
        }
        return i;
    }
}
\end{lstlisting}
\begin{lstlisting}[style=Java, float, floatplacement=H, caption={Excerpt of disassembled bytecode from code in \autoref{lst:commons-codec-util-original-src}.}, label={lst:commons-codec-util-bytecode}] 
class org.apache.commons.codec.net.Utils {
  static int digit16(byte) throws org.apache.commons.codec.DecoderException;
         0: ILOAD_0             //Parameter byte b
         1: I2C
         2: BIPUSH        16
         4: INVOKESTATIC  #19   //Character.digit:(CI)I            
         7: ISTORE_1            //Variable int i
         8: ILOAD_1
         9: ICONST_m1
        10: IF_ICMPNE     37
//org/apache/commons/codec/DecoderException
        13: NEW           #17
        16: DUP
        17: NEW           #25   //java/lang/StringBuilder
        20: DUP
//"Invalid URL encoding: not a valid digit (radix 16):"
        21: LDC           #27
//StringBuilder."<init>":(Ljava/lang/String;)V
        23: INVOKESPECIAL #29
        26: ILOAD_0
//StringBuilder.append:(I)Ljava/lang/StringBuilder;
        27: INVOKEVIRTUAL #32
//StringBuilder.toString:()Ljava/lang/String;
        30: INVOKEVIRTUAL #36
//DecoderException."<init>":(Ljava/lang/String;)V
        33: INVOKESPECIAL #40
        36: ATHROW
        37: ILOAD_1
        38: IRETURN
}
\end{lstlisting}

As mentioned, the key challenge of decompilation resides in the many ways in which information is lost during compilation. Consequently, Java decompilers need to make several assumptions when interpreting bytecode instructions, which can also be generated in different ways. To illustrate this phenomenon, \autoref{lst:commons-codec-util-fernflower-src} and \autoref{lst:commons-codec-util-dava-src} show the Java sources produced by the \fernflower and \dava decompilers when interpreting the bytecode of \autoref{lst:commons-codec-util-bytecode}. In both cases, the decompilation produces correct Java code (\ie, recompilable) with the same functionality than the input bytecode. Notice that \fernflower guesses that the series of \texttt{StringBuilder} (bytecode instruction $23$ to $27$) calls is the compiler's way of translating string concatenation and is able to revert it. On the contrary, the \dava decompiler does not reverse this transformation. As we can notice, the decompiled sources are different from the original in at least three points: 

\begin{lstlisting}[style=Java, float, floatplacement=H, caption={Decompilation result of \autoref{lst:commons-codec-util-bytecode} with \fernflower.}, label={lst:commons-codec-util-fernflower-src}] 
class Utils {
   private static final int RADIX = 16;
   static int digit16(byte b) throws DecoderException {
      int i = Character.digit((char)b, 16);
      if(i == -1) {
         throw new DecoderException("Invalid URL encoding: not a valid digit (radix 16): " + b);
      } else {
         return i;
      }
   }
}
\end{lstlisting}
\begin{lstlisting}[style=Java, float, floatplacement=H, caption={Decompilation result of \autoref{lst:commons-codec-util-bytecode} with \dava.}, label={lst:commons-codec-util-dava-src}] 
class Utils
{
    static int digit16(byte b)
        throws DecoderException
    {
        int i = Character.digit((char)b, 16);
        if(i == -1)
            throw new DecoderException((new StringBuilder()).append("Invalid URL encoding: not a valid digit (radix 16): ").append(b).toString());
        else
            return i;
    }
    private static final int RADIX = 16;
}
\end{lstlisting}

\begin{itemize}
    \item In the original sources, the local variable $i$ was \texttt{final}, but \javac lost this information during compilation.
    
    \item The \texttt{if} statement had originally no \texttt{else} clause. Indeed, when an exception is thrown in a method that do not catch it, the execution of the method is interrupted. Therefore, leaving the \texttt{return} statement outside of the \texttt{if} is equivalent to putting it inside an \texttt{else} clause.

    \item In the original code the String \texttt{"Invalid URL encoding: not a valid digit (radix 16): "} was actually computed with \texttt{"Invalid URL encoding: not a valid digit (radix " + URLCodec.RADIX + "): "}. In this case, \texttt{URLCodec.RADIX} is actually a final static field that always contains the value $16$ and cannot be changed. Thus it is safe for the compiler to perform this optimization, but the information is lost in the bytecode.
\end{itemize}

Besides, this does not include the different formatting choices made by the decompilers such as new lines placement and brackets usage for single instructions such as \texttt{if} and \texttt{else}.

\section{Methodology}\label{sec:methodology}

In this section, we introduce definitions, metrics and research questions. Next, we  detail the framework to compare decompilers and we describe the Java projects that form the set of case studies for this work.

\subsection{Definitions and Metrics}
\label{sec:def}

The value of the results produced by decompilation varies greatly depending on the intended use of the generated source code. In this work, we evaluate the decompilers capacity to produce a faithful retranscription of the original sources. Therefore, we collect the following metrics.

\begin{defn}
\label{met:syntactic-c}
\textbf{Syntactic correctness.} The output of a decompiler is syntactically correct if it contains a valid Java program, \ie a Java program that is recompilable with a Java compiler without any error.
\end{defn}

When a bytecode decompiler generates source code that can be recompiled, this source code can still be  syntactically different from the original. 
We introduce a metric to measure the scale of such a difference according to the abstract syntax tree (AST) dissimilarity\cite{Falleri2014} between the original and the decompiled results. This metric, called \textit{syntactic distortion}, allows to measure the differences that goes beyond variable names. The description of the metric is as follows:

\begin{defn}
\label{def:syntactic-distortion}
\textbf{Syntactic distortion.} Minimum number of atomic edits required to transform the AST of the original source code of a program into the AST of the corresponding decompiled version of it. 
\end{defn}

In the general case, determining if two program are semantically equivalent is undecidable. For some cases, the decompiled sources can be recompiled into bytecode that is equivalent to the original, modulo reordering of the constant pool. We call these cases \textit{strictly equivalent} programs. We measure this equivalence with a bytecode comparison tool named Jardiff.\footnote{\url{https://github.com/scala/jardiff}}

Inspired by the work of~\cite{Le2014} and~\cite{Yanghunting2019}, we check if the decompiled and recompiled program is semantically equivalent modulo inputs. This means that for a given set of inputs, the two program produce equivalent outputs. In our case, we select the set of relevant inputs and assess equivalence based on the existing test suite of the original program.

\begin{defn}
\label{met:semi}
\textbf{Semantic equivalence modulo inputs.} We call a decompiled program semantically equivalent modulo inputs to the original if it passes the set of tests from the original test suite.
\end{defn}

In the case where the decompiled and recompiled program produce non-equivalent outputs, that demonstrates that the sources generated by the decompiler express a different behavior than the original. As explained by Hamilton and colleagues\cite{Hamilton2009}, this is particularly problematic as it can mislead decompiler users in their attempt to understand the original behavior of the program. We refer to theses cases as \textit{deceptive decompilation} results.

\begin{defn}
\label{def:deceptive}
\textbf{Deceptive decompilation:} Decompiler output that is syntactically correct but not semantically equivalent to the original input.
\end{defn}

\subsection{Research Questions}

We elaborated five research questions to guide our study on the characteristics of modern Java decompilers. 

\newcommand{\RQone}{To what extent is decompiled Java code syntactically correct?}
\newcommand{\RQtwo}{To what extent is decompiled Java code semantically equivalent modulo inputs?}

\newcommand{\RQthree}{To what extent do decompilers produce deceptive decompilation results?}
\newcommand{\RQfour}{What is the syntactic distortion of decompiled code?}
\newcommand{\RQfive}{To what extent the behavioral diversity of decompilers can be leveraged to improve the decompilation of Java bytecode?}

\textit{\textbf{RQ1: \RQone}}

In this research question, we investigate the effectiveness of decompilers for producing syntactically correct and hence recompilable source code from bytecode produced by the \javac and \ecj compilers.

\textit{\textbf{RQ2: \RQtwo}}

In this research question, we investigate on the semantic differences between the original source code and the outputs of the decompilers.

\textit{\textbf{RQ3: \RQthree}}

Le and colleagues\cite{Le2014} propose to use equivalence modulo inputs assessment as a way to test transformations that are meant to be semantic preserving (in particular compilation).
In this research question, we adapt this concept in the context  of decompilation testing. In this paper we rely on the existing test suite instead of generating inputs.

\textit{\textbf{RQ4: \RQfour}}
Even if decompiled bytecode is ensured to be syntactically and semantically correct, syntactic differences may remain as an issue when the purpose of decompilation is human understanding. Keeping the decompiled source code free of syntactic distortions is essential during program comprehension, as many decompilers can produce human unreadable code structures. In this research question, we compare the syntactic distortions produced by decompilers.

\textit{\textbf{RQ5: \RQfive}} 
As we observed during their comparison, each decompiler have their pros and cons. We evaluate if this diversity of features can be leveraged to boost the overall decompilation results. 

\subsection{Study Protocol}

\begin{figure}[t]
	\centering
	\includegraphics[origin=c,width=0.48\textwidth]{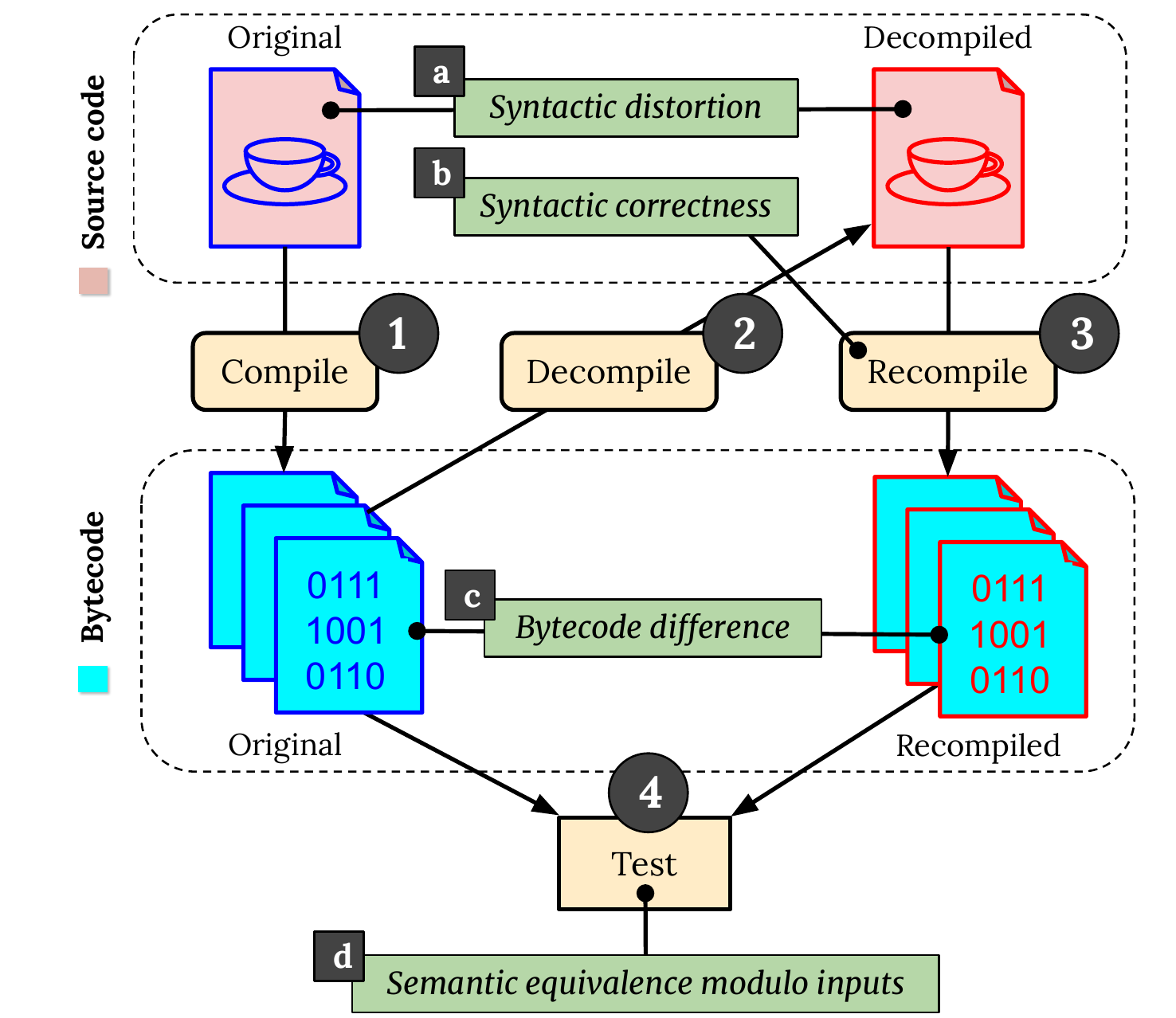}
	\caption{Java decompiler assessment pipeline with four evaluation layers: syntactic distortion, bytecode difference, syntactic correctness, and semantic equivalence modulo input.}
	\label{fig:pipeline}
\end{figure}

\autoref{fig:pipeline} represents the pipeline of operations conducted on every Java source file in our dataset. For each triplet \textit{<decompiler, compiler, project>}, we perform the following: 
\begin{enumerate}
    \item Compile the source files with a given compiler
    \item Decompile each class file  with a decompiler (there might be several classes if the source defines internal classes). If the decompiler does not return any error, we mark the source file as decompilable. Then, (a) we measure syntactic distortion by comparing the AST of the original source with the AST of the decompiled source.
    \item Recompile the class files with the given compiler. If the compilation is successful, we know that the decompiler produces (b) syntactically correct code. Then, we measure (c) the difference between the original  and the recompiled bytecode.
    \item Run the test cases on the recompiled bytecode. If the tests are successful, we mark the source as \textit{passTests} for the given triplet, showing that the decompiler produces   (d) semantically equivalent code modulo inputs.
\end{enumerate}

If one of these steps fails we do not perform the following steps and consider all the resulting metrics as not available. As decompilation can sometime produce  a program that does not stop, we set a $20$ minutes  timeout  on the test execution (the original test suites run under a minute on the hardware used for this experiment, a Core i5-6600K with 16Go of RAM).

The tests used to assess the semantic equivalence modulo inputs are those of the original project that cover the given Java file.\footnote{Coverage was assessed using yajta \url{https://github.com/castor-software/yajta}} We manually excluded the tests that fail on the original project (either flaky or because versioning issue). The list of excluded tests is available as part of our experiments.

\subsection{Study Subjects}

\textbf{Decompilers.}
\autoref{tab:decompilers} shows the set of decompilers under study. We have selected Java decompilers that are (i) freely available, and (ii) have been active in the last two years. We add \jode in order to compare our results with a legacy decompiler, and because the previous survey by Hamilton and colleagues'  considers it to be  one of the best decompilers \cite{Hamilton2009}.

The column \textsc{Version} shows the version used (some decompilers do not follow any versioning scheme). We choose the latest release if one exist, if note the last commit available the 09-05-2019. The column \textsc{\#Commits} represents the number of commits in the decompiler project, in cases where the decompiler is a submodule of a bigger project (\eg, \dava and \fernflower) we count only commits affecting the submodule. The column \textsc{\#LOC} is the number of line of code in all Java files (and Python files for \krakatau) of the decompiler, including sources, test sources and resources counted with \textit{cloc}.\footnote{\url{http://cloc.sourceforge.net/}}

\begin{table}[t!]
\caption{Characteristics of the Studied Decompilers}
\centering
\scriptsize
\begin{tabular}{lcccc}
			\hline
			\textsc{\textbf{Decompiler}}  & \textsc{\textbf{Version}} & \textsc{\textbf{Status}} & \textsc{\textbf{\#Commits}} & \textsc{\textbf{\#LOC}} \\
			\hline
    		\cfr \cite{cfr}   & $0.141$ & Active & $1433$ & $52098$ \\
    		\dava \cite{dava}        & $3.3.0$ & Updated 2018-06-15 & $14$ & $22884$ \\
    		\fernflower \cite{fernflower}   & NA* & Active & $453$ & $52118$ \\
    		\jadx \cite{jadx} & 0.9.0 & Active & $970$ & $55335$ \\
    		\jd \cite{jdcore} & 1.0.0 & Active & NA** & $36730$ \\
    		\jode \cite{jode} & 1.1.2-pre1 & Updated 2004-02-25 & NA** & $30161$ \\
    		\krakatau \cite{krakatau}  & NA* & Updated 2018-05-13 & $512$ & $11301$ \\
    		\procyon \cite{procyon}  & $0.5.34$ & Active & $1080$ & $122147$ \\
            \hline
		\end{tabular}
	
	\label{tab:decompilers}
	\begin{tablenotes}
      \footnotesize
      \centering
      \item * Not following any versioning scheme.
      \item ** CVS not available at the date of the present study.
    \end{tablenotes}
\end{table}

\textbf{Projects.}
In order to get a set of real world Java projects to evaluate the eight  decompilers, we reuse the set of projects of Pawlak and colleagues\cite{Pawlak2015}. To these $13$ projects we added a fourteenth one named \texttt{DcTest} made out of examples collected from previous decompiler evaluations \cite{Hamilton2009,Kostelansky2017}.\footnote{\url{http://www.program-transformation.org/Transform/JavaDecompilerTests}} \autoref{tab:dataset} shows a summary of this dataset: the Java version in which they are written, the number of Java source files, the number of unit tests as reported by Apache Maven, and the number of Java lines of code in their sources.

As different java compilers may translate the same sources into different bytecode representations,\footnote{\url{https://www.benf.org/other/cfr/eclipse-differences.html}} we employed the two most used Java compilers: \javac and \ecj (we use versions $1.8.0\_17$ and $13.13.100$, respectively). We compiled all the $14$ projects with both compilers (except \texttt{commons-lang} that we failed to build it with \ecj). This represents $1887$ class files for each compiler that we use to evaluate syntactic correctness of decompiler outputs in RQ1 and syntactic distortion in RQ4. 
We select only those that contain code executed by tests ($2397$ grouping files generated by the two compilers) to evaluate semantic correctness in RQ2 and RQ3. 

\begin{table}[t!]
\caption{Characteristics of the projects used to evaluate decompilers}
\scriptsize
\centering
    \begin{tabular}{lcccc}
		\hline
		\textsc{\textbf{Project name}} & \textsc{\textbf{Java version}} & \textsc{\textbf{\#Classes}} & \textsc{\textbf{\#Tests}} & \textsc{\textbf{\#LOC}}\\
		\hline
        Bukkit	& $1.6$ & $642$ & $906$ & $60800$\\
        Commons-codec & $1.6$ & $59$ & $644$ & $15087$\\
        Commons-collections	& $1.5$ & $301$ & $15067$ & $62077$\\
        Commons-imaging	& $1.5$ & $329$ & $94$ & $47396$\\
        Commons-lang & $1.8$ & $154$ & $2581$ & $79509$\\
        DiskLruCache & $1.5$ & $3$ & $61$ & $1206$\\
        JavaPoet*	& $1.6$ & $2$ & $60$ & $934$\\
        Joda time	& $1.5$ & $165$ & $4133$ & $70027$\\
        Jsoup	& $1.5$ & $54$ & $430$ & $14801$\\
        JUnit4	& $1.5$ & $195$ & $867$ & $17167$\\
        Mimecraft	& $1.6$ & $4$ & $14$ & $523$\\
        Scribe Java	& $1.5$ & $89$ & $99$ & $4294$\\
        Spark & $1.8$ & $34$ & $54$ & $4089$\\
        DcTest** & $1.5-1.8$ & $10$ & $9$ & $211$\\
        \hline
        \textsc{\textbf{Total}} &  & \textbf{$2041$}  & \textbf{$25019$} & \textbf{$378121$}\\
        \hline
	\end{tabular}

	\label{tab:dataset}
	\begin{tablenotes}
	\centering
      \footnotesize
      \item (*) Formerly named JavaWriter.
      \item (**) Examples collected from previous decompilers evaluation.
    \end{tablenotes}
\end{table}
\section{Experimental Results}\label{sec:results}

\subsection{\textbf{RQ1: (syntactic correctness)} \RQone}
This research question  investigates to what extent the source code produced by the different decompilers is syntactically correct, meaning that the decompiled code compiles. We also investigate the effect of the compiler that produces the bytecode on the decompilation results.

\begin{figure}[htb]
	\centering
	\includegraphics[origin=c,width=0.485\textwidth]{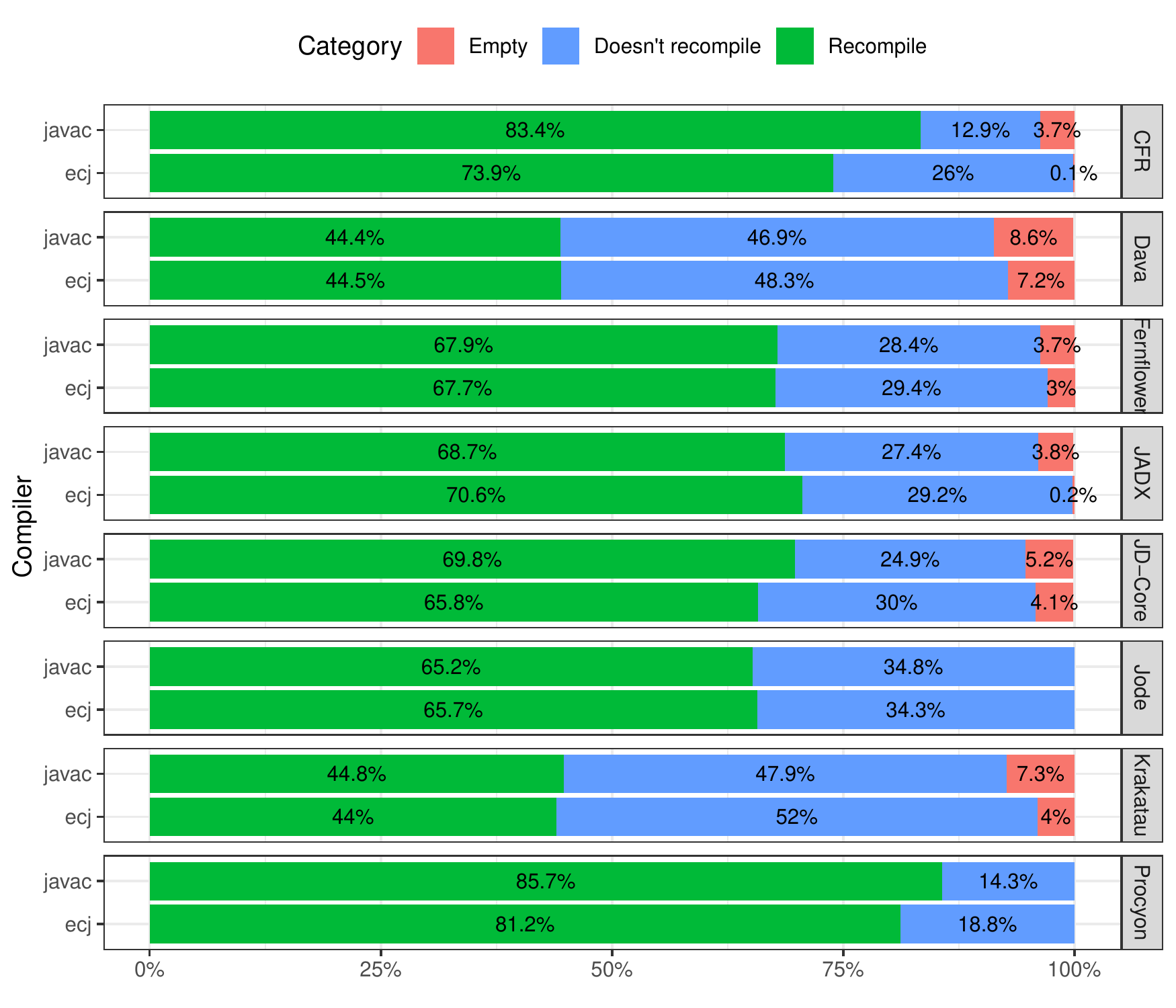}
	\caption{Successful recompilation ratio after decompilation for all considered decompilers.}
	\label{fig:decompilation_categories_results}
\end{figure}

\autoref{fig:decompilation_categories_results} shows the ratio of decompiled classes that are syntactically correct per pair of compiler and decompiler. The horizontal axis shows the ratio of syntactically correct output in green, the ratio of syntactically incorrect output in blue, and the ratio of empty output in red (an empty output occurs, e.g. when the decompiler crashes). The vertical axis shows the compiler on the left and decompiler on the right. For example, \procyon, shown in the last row, is able to produce a syntactically correct source code for $1609$ ($85.3\%$) class files compiled with \javac, and produce a non empty syntactically incorrect output for $278$ ($14.7\%$) of them. On the other hand, when sources are compiled with \ecj, \procyon generates syntactically correct sources for $1532$ ($82.2\%$) of the class files and syntactically incorrect for $355$ ($18.8\%$) sources. In other words, \procyon is slightly more effective when used against code compiled with \javac. 
It is interesting to notice that not all decompiler authors have decided to handle error the same way. Both \procyon and \jode's developers have decided to always return source files, even if incomplete (for our dataset). Additionally, when \cfr and \procyon detect a method that they cannot decompile properly, they may replace the body of the method  by a single \texttt{throw} statement and comment explaining the error. This leads to syntactically correct code, but not semantically equivalent.

The ratio of syntactically correct decompiled code ranges from $85.7\%$ for \procyon on \javac inputs (the best), down to $44\%$ for \krakatau on \ecj (the worst). Overall, no decompiler is capable of correctly handling the complete dataset. This illustrates the challenges of Java bytecode decompilation, even for bytecode that has not been obfuscated, as in the case of our experiments.

We note that syntactically incorrect decompilation can still be useful for reverse engineering. However, an empty output is useless: the ratio of class files for which the decompilation completely fails is never higher than $8.6\%$ for \dava on \javac bytecode.


Intuitively, it seems that the compiler has an impact on decompilation effectiveness. To verify this, we use a $\chi^{2}$ test on the ratio of classfile decompiled into syntactically correct source code depending on the used compiler, \javac versus \ecj.
The compiler variable has an impact for three decompilers and no impact for the remaining five at $99\%$ confidence level.
The test rejects that the compiler has no impact on the decompilation syntactic correctness ratio for  \cfr, \procyon and \jd (p-value $10^{-14}$, $0.00027$ and $0.006444$). For the five other decompilers we do not observe a significant difference between \javac and \ecj (p-values: \dava $0.15$, \fernflower $0.47$, \jadx $0.17$, \jode $0.50$, and \krakatau $0.09$). Note that beyond syntactic correctness, the compiler may impact the correctness of the decompiled code, this will be discussed in more details in Section~\ref{sec:rq3-results}.

To sum up, \procyon and \cfr are the decompilers that score the highest on syntactic correctness. The three decompilers ranking the lowest are \jode, \krakatau and \dava. It is interesting to note that those three are no longer actively maintained.
\pagebreak
\begin{mdframed}[style=mpdframe]
\textbf{Answer to RQ1:} 
No single decompiler is able to produce syntactically correct sources for more than $85.7\%$ of class files in our dataset. The implication for decompiler users is that decompilation of Java bytecode cannot be blindly applied and do require some additional manual effort. Only few cases make all decompiler fail, which suggest that using several decompilers in conjunction could help to achieve better results.
\end{mdframed}

\subsection{\textbf{RQ2: (semantic equivalence)} \RQtwo}

To answer this research question, we  focus on the $2397$ class files that are covered by at least one test case. When decompilers produce sources that compile, we investigate the semantic equivalence of the decompiled source and their original. To do so, we split recompilable outputs in three categories: (i) \textit{semantically equivalent:} the code is recompiled into bytecode that is strictly identical to the original (modulo reordering of constant pool, as explained in Section \ref{sec:def}), (ii) \textit{semantically equivalent modulo inputs:} the output is recompilable and passes the original project test suite (i.e. we cannot prove that the decompiled code is semantically different), and (iii) \textit{semantically different:} the output is recompilable but it does not pass the original test suite (deceptive decompilation, as explained in Definition \autoref{def:deceptive}).

\begin{lstlisting}[style=Java, float, floatplacement=H, caption={Excerpt of org/apache/commons/codec/language/bm/Lang.class bytecode compiled with \javac and decompiled with \cfr:
    Lines in red are in the original byte code, while lines in green are from the recompiled sources.}, label={lst:lang-bytecode}]
     `IFEQ L2`
     @IFNE L2@
     @GOTO L0@
   @L2@
     ALOAD 5
     INVOKESTATIC Lang$LangRule.access$100 (LLang$LangRule;)Z
     IFEQ L3
     ALOAD 3
     ALOAD 5
     INVOKESTATIC Lang$LangRule.access$200 (Lang$LangRule;)Ljava/util/Set;
     `INVOKEINTERFACE Set.retainAll (LCollection;)Z` 
     @INVOKEVIRTUAL HashSet.retainAll (LCollection;)Z@
     (itf)
     POP
     `GOTO L2`
     @GOTO L0@
\end{lstlisting}

Let us first discuss an interesting example of semantic equivalence of decompiled code. \autoref{lst:lang-bytecode} shows an example of bytecode that is different when decompiled-recompiled but equivalent modulo inputs to the original. Indeed, we can spot two differences: the control flow blocks are not written in the same order (\texttt{L2} becomes \texttt{L0}) and the condition evaluated is reversed (\texttt{IFEQ} becomes \texttt{IFNEQ}), which leads to an equivalent control flow graph. The second difference is that the type of a variable originally typed as a \texttt{Set} and instantiated with an \texttt{HashSet} has been transformed into a variable typed as an \texttt{HashSet}, hence once \texttt{remainAll} is invoked on the variable \texttt{INVOKEINTERFACE} becomes directly \texttt{INVOKEVIRTUAL}. This is still equivalent code. 

\begin{figure}[t]
	\centering
	\includegraphics[origin=c,width=0.47\textwidth]{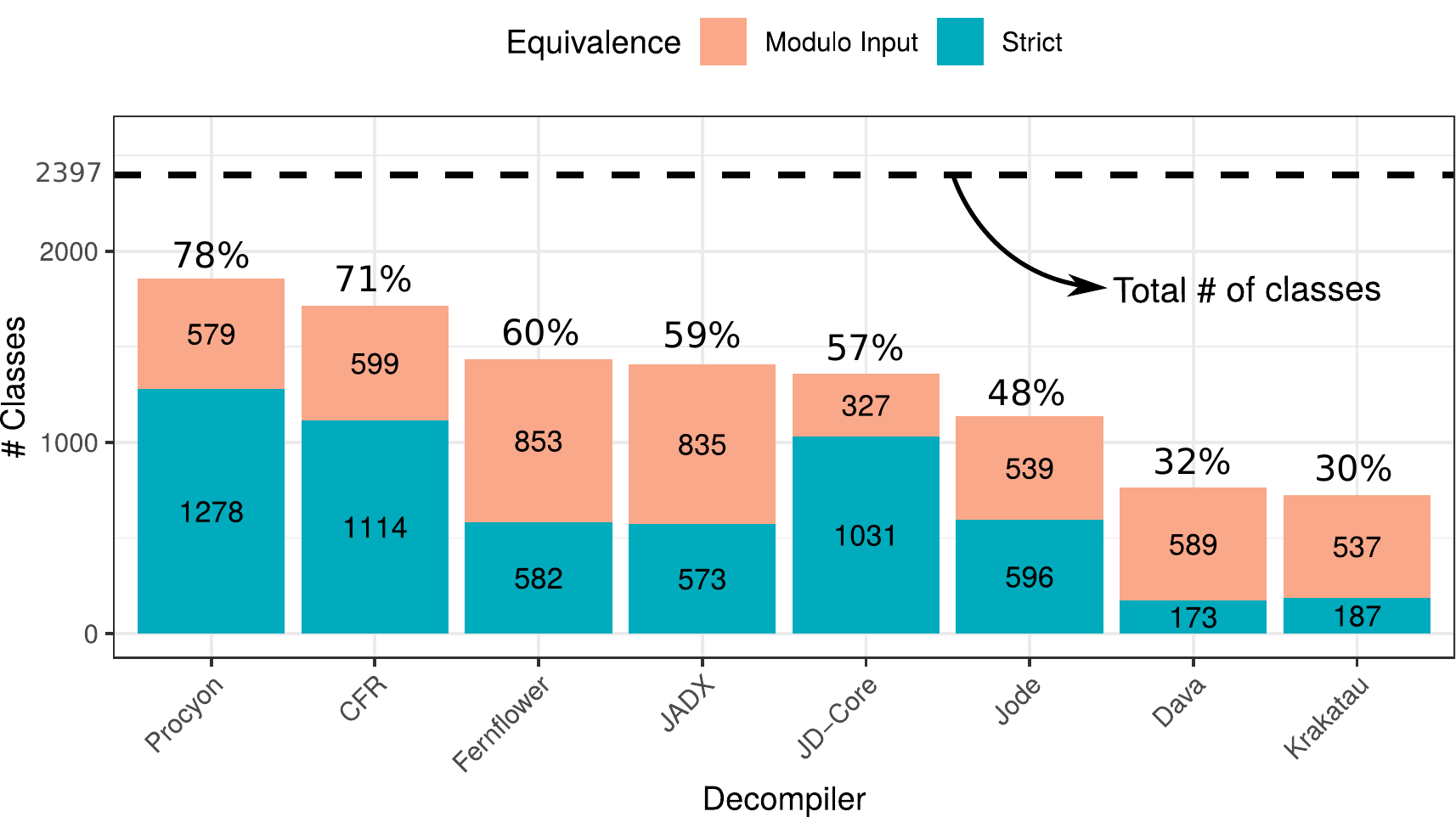}
	\caption{Equivalence results for each decompiler on all the classes of the studied projects covered by at least one test.}
	\label{fig:equivalence_categories_results}
\end{figure}

Now we discuss the results globally. \autoref{fig:equivalence_categories_results} shows the recompilation outcomes of decompilation regarding semantic equivalence for the $2397$ classes under study.
The horizontal axis shows the eight different decompilers. The vertical axis shows the number of classes decompiled successfully.
Strictly equivalent output are shown in blue, equivalent classes modulo input classes are shown in orange. For example, \cfr (second bar) is able to decompile correctly $1713$ out of $2397$ classes ($71\%$), including $1114$ classes that are recompilable into strictly equivalent bytecode, and $599$ that are recompilable into equivalent bytecode modulo inputs. 

The three decompilers that are not actively maintained anymore (\jode, \dava and \krakatau) handle less than $50\%$ of the cases correctly (recompilable and pass tests). On the other hand, \procyon and \cfr have the highest ratio of equivalence modulo inputs of $78\%$ and $71\%$, respectively. 

\begin{mdframed}[style=mpdframe]
\textbf{Answer to RQ2:} The number of classes for which the decompiler produces EMI semantically equivalent varies a lot from one decompiler to another. The source code generated by the decompilers is usually not strictly identical to the original, still many of the decompiled classes are semantically equivalent modulo inputs. 
For end users, it means that the state of the art of Java decompilation does not guarantee semantically correct decompilation, and care must be taken not to blindly trust in the decompiled code.
\looseness=-1
\end{mdframed}

\subsection{\textbf{RQ3: (bug finding)} \RQthree}
\label{sec:rq3-results}
As explained by Hamilton and colleagues~\cite{Hamilton2009}, while a syntactically incorrect decompilation output may still be useful to the user, syntactically correct but semantically different output is more problematic. Indeed, this may mislead the user by making her believe in a different behavior than the original program. We call this case \textit{deceptive decompilation} (as explained in  Definition \autoref{def:deceptive}). When such cases occur, since the decompiler produces an output that is semantically different from what is expected, they may be considered as decompilation bugs.

\begin{figure}[t]
	\centering
	\includegraphics[origin=c,width=0.475\textwidth]{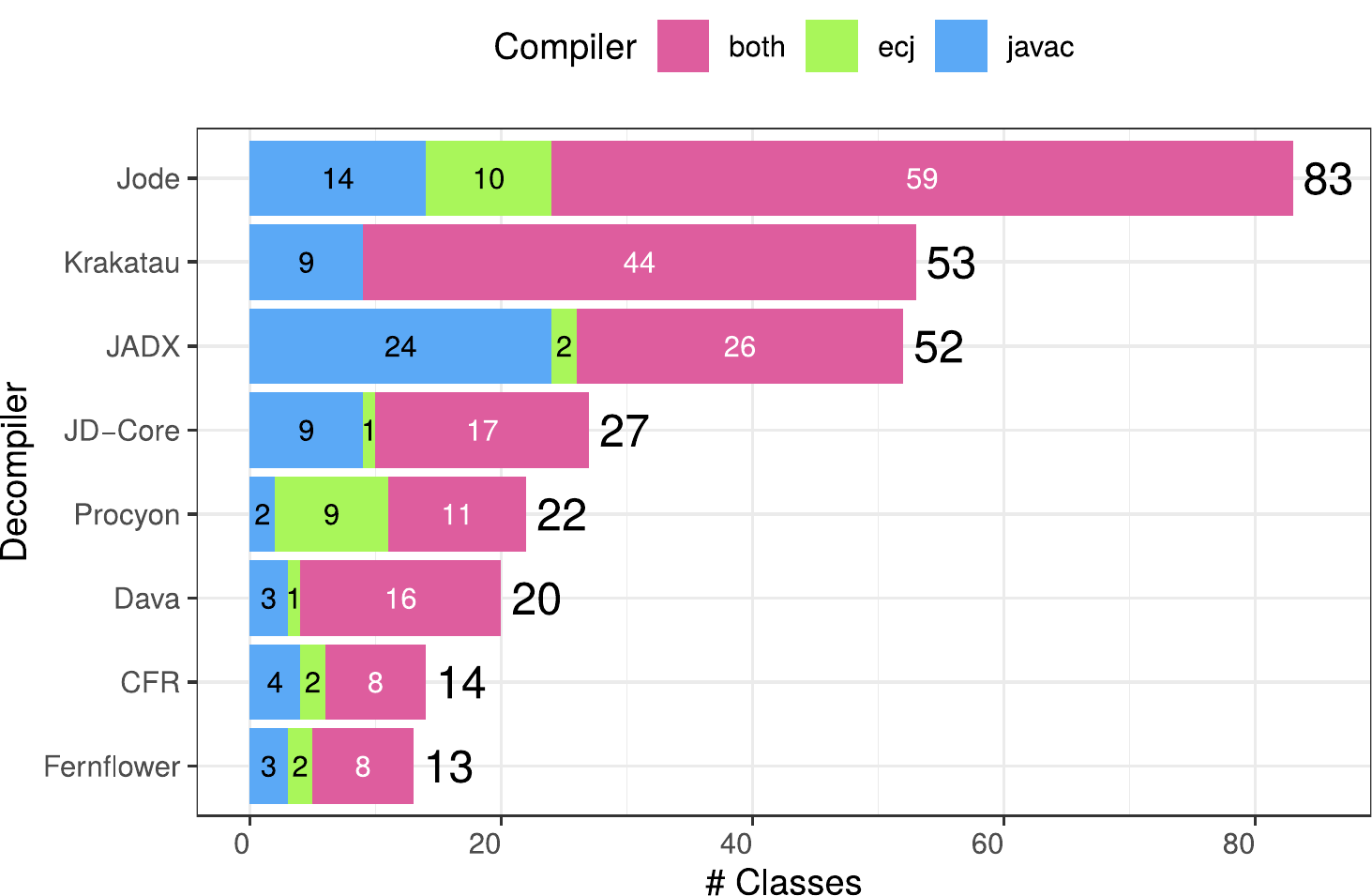}
	\caption{Deceptive decompilations per decompiler.}
	\label{fig:r_not_t_compilers}
\end{figure}

\autoref{fig:r_not_t_compilers} shows the distribution of bytecode classes that are deceptively decompiled. Each horizontal bar groups deceptive decompilation per decompiler. The color indicates which compiler was used to produce the class file triggering the error. In blue is the number of classes leading to a decompilation error only when compiled with \javac, in green only when compiled with \ecj, and in pink it is the number of classes triggering a decompilation error with both compilers. The sum of these classes is indicated by the total on the right side of each bar.
Note that the bars in \autoref{fig:r_not_t_compilers} represents the number of bug manifestations, which are not necessarily distinct bugs: the same decompiler bug can be triggered by different class files from our benchmark.

Overall, \jode is the least reliable decompiler, with 83 decompilation bug instances in our benchmark. While \fernflower produces the least deceptive decompilations on our benchmark ($13$), it is interesting to note that \cfr produces only one more deceptive decompilation ($14)$ but that correspond to less bugs per successful decompilation. This makes \cfr the most reliable decompiler on our benchmark.

We manually inspected $10$ of these bug manifestations.
$2$ of them were already reported by other users.
We reported the other $8$ of them to the authors of decompilers.\footnote{\url{https://github.com/castor-software/decompilercmp/tree/master/funfacts}} The sources of errors include incorrect cast operation, incorrect control-flow restitution, auto unboxing errors, and incorrect reference resolution. Below we detail two of these bugs.

\subsubsection{Case study: incorrect reference resolution}
We analyze the class \texttt{org.bukkit.Bukkit} from the Bukkit project. An excerpt of the original Java source code is given in \autoref{lst:bukkit-original-decompiled}. The method \texttt{setServer} implements a setter of the static field \texttt{Bukkit.server}. This is an implementation of the common Singleton design pattern.
In the context of method \texttt{setServer}, \texttt{server} refers to the parameter of the method, while \texttt{Bukkit.server} refers to the static field of the class \texttt{Bukkit}.

\begin{lstlisting}[style=Java, float, floatplacement=H, caption={Exerpt of differences in org/bukkit/Bukkit.java:63 original (in red) and decompiled with \jadx sources (in green).}, label={lst:bukkit-original-decompiled}] 
public final class Bukkit {
    private static Server server;
	[...]
    public static void setServer(Server server) {
        `if (Bukkit.server != null) {`
        @if (server != null) {@
            throw new UnsupportedOperationException(
            "Cannot redefine singleton Server");
        }
        `Bukkit.server = server;`
        @server = server;@
        [...]
    }
\end{lstlisting}

When this source file is compiled with \textit{javac}, it produces a file \texttt{org/bukkit/Bukkit.class} containing the bytecode translation of the original source.
\autoref{lst:bukkit-bytecode} shows an excerpt of this bytecode corresponding to the \texttt{setServer} method (including lines are filled in red, while excluding lines are filled in green)

\begin{lstlisting}[style=Java, float, floatplacement=H, caption={Exerpt of org/bukkit/Bukkit.class bytecode compiled with \javac: Lines in red are in the original byte code, while lines in green are from the recompiled sources (decompiled with \jadx).}, label={lst:bukkit-bytecode}] 
public static setServer(Lorg/bukkit/Server;)V
     `GETSTATIC org/bukkit/Bukkit.server :`
     `Lorg/bukkit/Server;`
     @ALOAD 0@
     IFNULL L0
     NEW java/lang/UnsupportedOperationException
     DUP
     ATHROW
    L0
     ALOAD 0
     `PUTSTATIC org/bukkit/Bukkit.server :`
     `Lorg/bukkit/Server;`
     @ASTORE 0@
     ALOAD 0
     INVOKEINTERFACE org/bukkit/Server.getLogger ()Ljava/util/logging/Logger; (itf)
     NEW java/lang/StringBuilder
\end{lstlisting}

When using the \jadx decompiler on \texttt{org/bukkit/Bukkit.class} it produces decompiled, with an excerpt shown in \autoref{lst:bukkit-original-decompiled}
In this example, the decompiled code is not semantically equivalent to the original version. Indeed, inside the \texttt{setServer} method the references to the static field \texttt{Bukkit.server} have been simplified into \texttt{server} which is incorrect in this scope as the parameter \texttt{server} overrides the local scope. In the bytecode of the recompiled version (\autoref{lst:bukkit-bytecode}, including lines are filled in green), we can observe  that instructions accessing and writing the static field (\texttt{GETSTATIC}, \texttt{PUTSTATIC}) have been replaced by instructions accessing and writing the local variable instead (\texttt{ALOAD}, \texttt{ASTORE}).

When the test suite of \textit{Bukkit} runs on the recompiled bytecode, the $11$ test cases covering this code fail, as the first access to \texttt{setServer} will throw an exception instead of normally initializing the static field \texttt{Bukkit.server}. This is clearly a bug in \jadx.

\subsubsection{Case study: Down cast error}

\begin{lstlisting}[style=Java, float, floatplacement=H, caption={Excerpt of differences in FastDatePrinter original (in red) and decompiled with \procyon sources (in green).}, label={lst:FastDatePrinter-procyon}]
    protected StringBuffer applyRules(final Calendar calendar, final StringBuffer buf) {
        `return (StringBuffer) applyRules(calendar,`
                                         `(Appendable) buf);`
        @return this.applyRules(calendar, buf);@
    }
    
    private <B extends Appendable> B applyRules(final Calendar calendar, final B buf) {...}
\end{lstlisting}

\autoref{lst:FastDatePrinter-procyon} illustrates the differences between the original sources of \texttt{org/apache/commons/lang3/time/FastDatePrinter} and the decompiled sources produced by \procyon. The line in red is part of the original, while the line in green is from the decompiled version.
In this example, method \texttt{applyRules} is overloaded, i.e. it has two implementations: one for a \texttt{StringBuffer} parameter and one for a generic \texttt{Appendable} parameter (\texttt{Appendable} is an interface that \texttt{StringBuffer} implements). The implementation for \texttt{StringBuffer}   down casts \texttt{buf} into \texttt{Appendable}, calls the method handling \texttt{Appendable} and casts the result back to \texttt{StringBuffer}. 
In a non ambiguous context, it is perfectly valid to call a method which takes  \texttt{Appendable} arguments on an instance of a class that implements that interface. But in this context, without the down cast to \texttt{Appendable}, the Java compiler will resolve the method call \texttt{applyRules} to the most concrete method. In this case, this will lead \texttt{applyRules} for \texttt{StringBuffer} to call itself instead of the other method. When executed this will lead to an infinite recursion ending in a StackOverflowError. Therefore, in this example, \procyon changes the behavior of the decompiled program   and introduces a bug in it.

\begin{mdframed}[style=mpdframe]
\textbf{Answer to RQ3:} 
Our empirical results indicate that no decompiler is free of deceptive decompilation bugs. 
The developers of decompilers may benefit from the equivalent modulo input concept to find bugs in the wild and extend their test base.
Two bugs found during our study have already been fixed by the decompiler authors, and three other have been acknowledged.

\looseness=-1
\end{mdframed}

\subsection{\textbf{RQ4: (ASTs difference)} \RQfour}

The quality of decompilation depends not only on its syntactic compilability and semantic equivalence but also on how well a human can understand the behavior of the decompiler program. The code produced by a decompiler may be syntactically and semantically correct but yet hard to read for a human.
In this research question, we evaluate how far the decompiled sources are from the original code.
We measure the syntactic distortion between the original and the decompiled sources as captured by AST differences (Definition \autoref{def:syntactic-distortion}). 

\begin{figure}[htb]
	\centering
	\includegraphics[origin=c,width=0.49\textwidth]{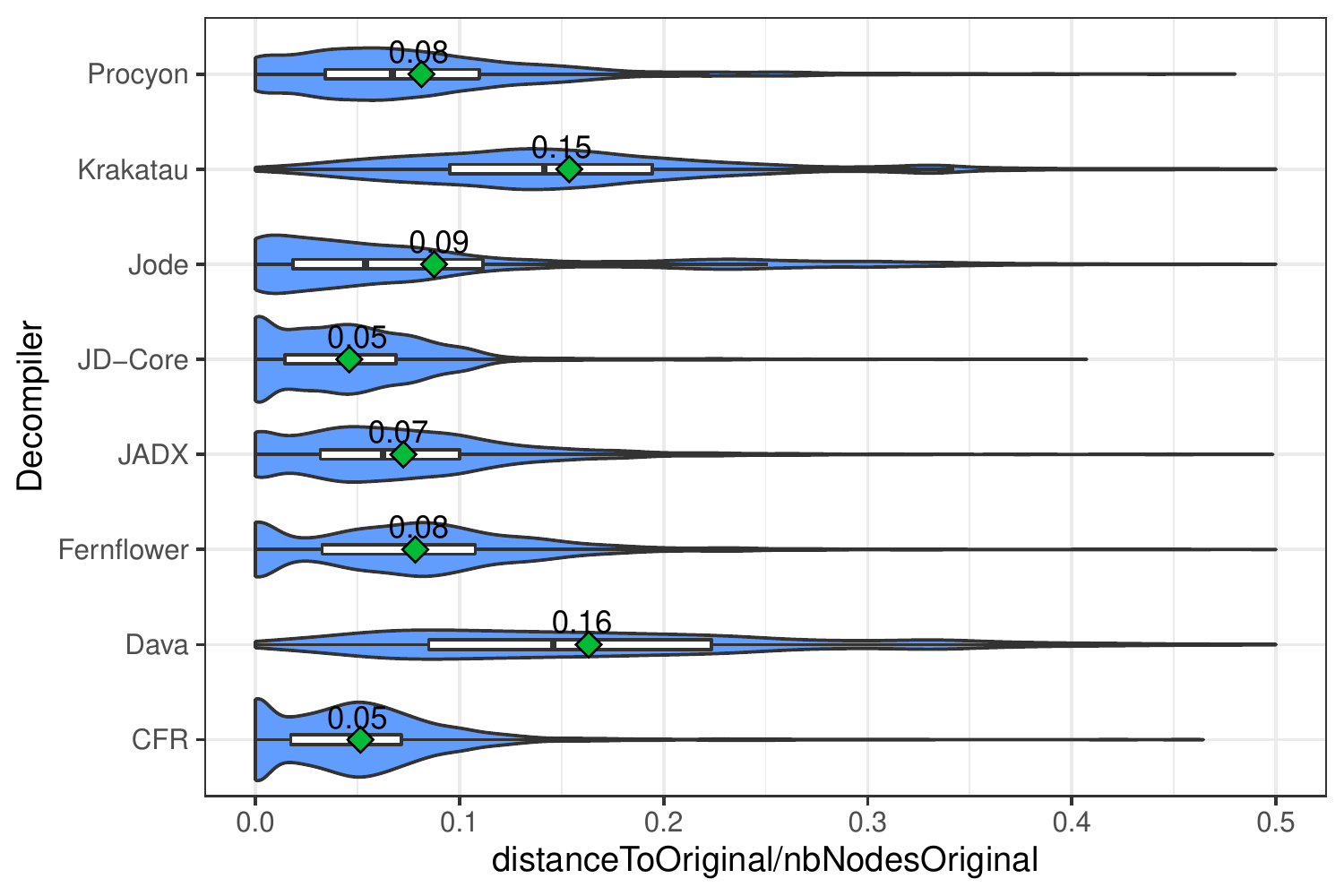}
	\caption{Distribution of ASTs differences between the original and the decompiled source code. Green diamonds indicate average.}
	\label{fig:distances}
	\vspace{-0.5cm}
\end{figure}




\autoref{fig:distances} shows  the distribution of syntactic distortion present in syntactically correct decompiled code, with one violin plot per decompiler. The green diamond marks the average syntactic distortion. For example, the syntactic distortion values of the \jode decompiler have a median of $0.05$, average of $0.09$, 1st-Q and 3rd-Q of $0.01$ and $0.11$, respectively. In this figure, lower is better: a lower syntactic distortion means that the decompiled sources are more similar to their original counterparts.

\cfr and \jd introduce the least syntactic distortion, with high proportion of cases with no syntactic distortion at all (as we exclude renaming). Their median and average syntactic distortion are close to $0.05$, which correspond to $5$ edits every $100$ nodes in the AST of the source program. On the other extreme, \dava and \krakatau introduce the most syntactic distortion with average of $16$ edit per $100$ nodes (and resp. $15$). They also have almost no cases for which they produce sources with no syntactic distortion. It is interesting to note that \dava  makes no assumption on the provenance of the bytecode\cite{Miecznikowski2002}. This partly explains the choice of its author to not reverse some of the optimization made by Java compilers (See example introduced in \autoref{sec:background}.).

\autoref{lst:foo-original-fernflower} shows the differences on the resulting source code after decompiling the \texttt{Foo} class from \texttt{DcTest} with \fernflower. As we can observe, both Java program represent a semantically equivalent program. Yet, their ASTs contain substantial differences. For this example, the edit distance is $3/104$ as it contains three tree edits: \texttt{MOVE} the return node, and \texttt{DELETE} the break node and the continue node (the original source's AST contained $104$ nodes). 

Note that some decompilers perform some transformations on the sources they produce on purpose to increase readability. Therefor, it is perfectly normal to observe some minimal syntactic distortion, even for decompilers producing readable sources. But as our benchmark is composed of non obfuscated sources, it is expected that a readable output will not fall too far from the original.

\textbf{}\begin{lstlisting}[style=Java, float, floatplacement=H, caption={Excerpt of differences in Foo original and decompiled with \fernflower sources.}, label={lst:foo-original-fernflower}]
public class Foo {
  public int foo(int i, int j) {
    while (true) {
      try {   
        while (i < j) i = j++ / i;
        @return j;@
      } catch (RuntimeException re) {
        i = 10;
        `continue;`
      }
      `break;`                   
    }
    `return j;`
  }
}
\end{lstlisting}

	\vspace{-0.5cm}
\begin{mdframed}[style=mpdframe]
\textbf{Answer to RQ4:} 
All decompilers present various degrees of  syntactic distortion between the original source code and the decompiled bytecode. This reveals that all decompilers adopt different strategies to craft source code from bytecode. Our results suggest that syntactic distortion can be used  by decompiler developers to improve the alignment between the decompiled sources and the original. Also, decompiler users can use this analysis when deciding which decompiler to employ.
\looseness=-1
\end{mdframed}

\subsection{\textbf{RQ5: (Multi-decompiler evaluation)} \RQfive}

In the previous research questions, we observe that different decompilers produce source code that varies in terms of syntactic correctness, semantic equivalence and syntactic distortion. As no decompiler can perfectly perform the decompilation task regarding all these aspects, developers may use several decompilers.\footnote{The website \url{http://www.javadecompilers.com} indeed proposes to leverage this multiplicity of decompilers.} In this section, we investigate what can be gained by joining the forces of multiple decompilers.

\begin{figure}[t]
	\centering
	\includegraphics[origin=c,width=0.42\textwidth]{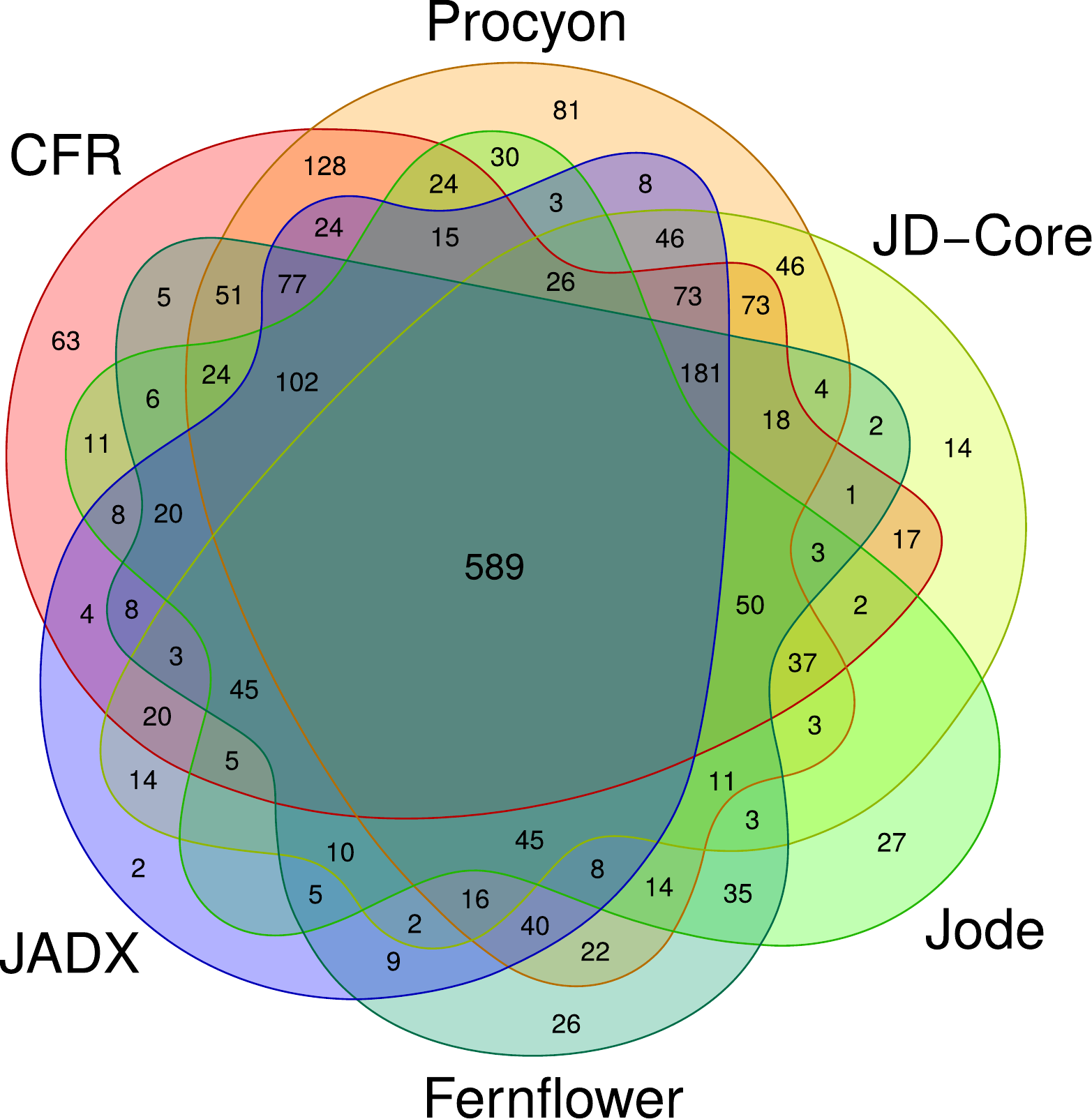}
	\caption{Venn diagram of syntactically and semantically equivalent modulo inputs decompilation results.}
	\label{fig:venn_diagram_recompilable}
	\vspace{-0.5cm}
\end{figure}

\autoref{fig:venn_diagram_recompilable} shows a Venn Diagram of syntactically and semantically equivalent classes modulo input for decompiled/recompiled classes. We exclude \dava and \krakatau because they that do not handle correctly any unique class file.
Indeed, 6/8 decompilers have cases for which they are the only decompiler able to handle it properly. These cases represent $276/2397$ classes. Only $589/2397$ classes are handled correctly by all of these 6 decompilers. Furthermore, $157/2397$ classes are not correctly handled by any of the considered decompilers.

To assess the benefit of using multiple decompilers instead of one, we have implemented a naive \meta that uses each decompiler one by one until it finds a syntactically correct decompilation result. The order of decompiler tried follows a ranking by decreasing success rate according to the six most successful decompilers in terms of \semi rates. 
In this manner, we can compare the effectiveness of this naive meta-decompiler with respect to the other decompilers taken in isolation.

\begin{table}[t!]
    \centering
    \scriptsize
    \caption{Summary results of the studied decompilers plus \meta}
    \begin{tabular}{lccccccc}
      \hline
     \textsc{\textbf{Decompiler}} & \textsc{\textbf{\#Recompilable}} & \textsc{\textbf{\#PassTest}} & \textsc{\textbf{\#Deceptive}} & \textsc{\textbf{ASTDiff}} \\ 
      \hline
      \cfr & $3097$ ($0.79$) & $1713$ ($0.71$) &  $22$ & $0.05$ \\ 
      \dava & $1747$ ($0.44$) & $762$ ($0.32$) &  $36$ & $0.17$ \\ 
      \fernflower & $2663$ ($0.68$) & $1435$ ($0.60$) &  $21$ & $0.08$ \\ 
      \jadx & $2736$ ($0.70$) & $1408$ ($0.59$) &  $78$ & $0.07$ \\ 
      \jd & $2726$ ($0.69$) & $1375$ ($0.57$) &  $82$ & $0.06$ \\ 
      \jode & $2569$ ($0.65$) & $1161$ ($0.48$) & $142$ & $0.09$ \\ 
      \krakatau & $1746$ ($0.44$) & $724$ ($0.30$) &  $97$ & $0.20$ \\ 
      \procyon & $3281$ ($0.84$) & $1869$ ($0.78$) &  $33$ & $0.08$ \\ 
      \meta & $3734$ ($0.95$) & $2174$ ($0.91$) &  $45$ & $0.08$ \\  
      \hline
    \end{tabular}
    \label{tab:overall}
	\vspace{-0.5cm}
\end{table}

\autoref{tab:overall} summarizes the quantitative results obtained from the previous research questions, and adds the effectiveness of the \meta as the last row. Each line corresponds to a decompiler. Column \texttt{\#Recompilable} shows the number of cases (and ratio) for which the decompiler produced a recompilable output; column \texttt{\#PassTest} shows the number of cases where the decompiled code passes those tests; column \texttt{\#Deceptive} indicate the number of cases that were recompilable but did not pass the test suite (i.e. a decompilation bug); column \texttt{\#ASTDist} indicate the average syntactic distortion among successfully decompiled cases.

Overall, the naive \meta implementation performs the best in terms of both syntactically correct and semantically equivalent modulo inputs criteria.
However, it is not the best in \texttt{\#Deceptive}, because it accumulates all bugs from \procyon and bugs from other decompilers that affect cases not handled by \procyon. 
Note that an user ready to give up performance could reorganize the order of decompilers tried by the \meta in order to optimize either \texttt{\#Deceptive} or \texttt{\#ASTDist} (with no impact on the number of syntactically correct cases). This shows that a decompiler user who would use the \meta  approach would obtain syntactically correct sources more frequently by $11$ points and semantically equivalent modulo inputs sources by $13$ points. 

As observed, the decompilation of Java is a non trivial task with no clear systematic solution. In order to produce useful results, decompiler developers make various assumptions about the source code that produced the bytecode, or about the compiler. For example \dava does not assume that the bytecode was produced from Java sources\cite{Miecznikowski2002}. \cfr does not trust information contained in the Local Variable Type Table,\footnote{\url{https://www.benf.org/other/cfr/faq.html}} as an obfuscation tool could change it without altering the behavior of the program. All these assumptions, and various strategies implemented by the  decompilers lead to a situation where the collection of implementations  successfully covers significantly more input cases that any individual implementation.

\begin{mdframed}[style=mpdframe]
\textbf{Answer to RQ5:} 
By leveraging the diversity of features present in existing decompilers, the naive \meta decompiler outperforms the other decompilers in terms of syntactic correctness (by $11$ percentage points compared to the best) and semantic equivalence modulo inputs (by $13$ percentage points). 
This quantitatively illustrates the benefit for decompiler users to try different decompiler instead of a single one. It also suggests research opportunities to approach the decompilation problem with a set of various decompilation strategies instead of a single one.\looseness=-1
\end{mdframed}

\section{Threats to Validity}\label{sec:threats}
In this section, we report about internal, external and reliability  threats against the validity of our results. 

\paragraph{Internal validity}

The internal threats are related to the metrics employed, especially those used to compare the syntactic distortion and semantic equivalence modulo inputs between the original and decompiled source code. Moreover, the coverage and quality of the test suite of the projects under study influences our observations about the semantic equivalence of the decompiled bytecode. To mitigate this threat, we select a set of mature open-source projects with good test suites as study subjects, and rely on state-of-the-art AST and bytecode differencing tools.

\paragraph{External validity}
The external threats refer to what extend the results obtained with the studied decompilers can be generalized to other Java projects. To mitigate this threat, we reuse an existing dataset of Java programs which we believe is representative of the Java world. Moreover, we added a handmade project which is a collection of classes used in previous decompilers evaluations as a baseline for further comparisons.\looseness=-1

\paragraph{Reliability validity} Our results are reproducible, the experimental pipeline presented in this study is publicly available online. We provide all necessary code to replicate our analysis, including AST metric calculations and statistical analysis via R notebooks.\footnote{\url{https://github.com/castor-software/decompilercmp/tree/master/notebooks}}\looseness=-1
\section{Related work}\label{sec:related}

This paper is related to previous works on bytecode analysis, decompilation and program transformations. In this section, we present the related work on Java bytecode decompilers along these lines.\looseness=-1

The evaluation of decompilers is closely related to the assessment of compilers. In particular, Le et al.~\cite{Le2014} introduce the concept of semantic equivalence modulo inputs to validate compilers by analyzing the interplay between dynamic execution on a subset of inputs and statically compiling a program to work on all kind of inputs. Naeem et al.~\cite{Naeem2007} propose a set of software quality metrics aimed at measuring the effectiveness of decompilers and obfuscators. In 2009, Hamilton et al.~\cite{Hamilton2009} show that decompilation is possible for Java, though not perfect. In 2017, Kostelansky et al.~\cite{Kostelansky2017} perform a similar study on updated decompilers. In 2018, Gusarovs~\cite{Gusarovs2018a} performed a study on five Java decompilers by analyzing their performance according to different handcrafted test cases. All those works demonstrate that fully Java bytecode decompilation is far from perfect.\looseness=-1

The objectives of decompilers are similar to disassemblers. However, instead of translating machine language into assembly language for different architectures~\cite{Vinciguerra2003, Khadra2016}, decompilers work at the high level of source code~\cite{Katz2019, Troshina2010, Emmerik2007}. Miecznikowski and Hendren~\cite{Miecznikowski2002} report about the problems and solutions found during the development of the Dava decompiler. They highlight particular issues related to expression evaluation on the Java stack, exceptions and synchronized blocks and type assignments. Disassemblers can sometimes be much more effective than decompilers, especially when the decompilation process goes wrong\cite{Khadra2016}.

Recently, Katz et al.~\cite{Katz2018} present a technique for decompiling binary code snippets using a model based on Recurrent Neural Networks, which produces source code that is more similar to human-written code and therefore more easy for humans to understand. This a remarkable attempt of driving decompilation to a specific goal. Schulte et al.~\cite{Schulte2018} use evolutionary search to improve and recombine a large population of candidate decompilations by applying source-to-source transformations gathered from a database of human-written sources. As an example of multi-tool that exploits diversity, Chen et al.~\cite{Chen2018} rely on various fuzzers to build an ensemble based fuzzer that gets better performance and generalization ability than that of any constituent fuzzer alone.

\section{Conclusion}\label{sec:conclusion}

Java bytecode decompilation is used for multiple purposes, ranging from reverse engineering to source recovery and understanding. 
In this work we proposed a fully automated pipeline to evaluate the Java bytecode decompilers' capacity to produce compilable, semantically equivalent and readable code. 
We proposed to use the concept of semantic equivalence modulo inputs to compare decompiled sources to their original counterpart. 
We applied this approach on $8$ available decompilers through a set of $2041$ classes from $14$ open-source projects compiled with $2$ different decompilers. 
The results of our analysis show that bytecode decompilation is a non trivial task that still requires human work. Indeed, even the highest ranking decompiler in this study  produces syntactically correct output for $84\%$ of classes of our dataset and semantically equivalent modulo inputs output for $78\%$. Meanwhile the diversity of implementation of these decompiler allow an user to combine several of them with significantly better results than by using a single one. 
In future work, we will explore the possibility to exploit this diversity of decompiler implementation automatically by merging the results of different decompilers via source code analysis and manipulations.

\newpage
\section*{Acknowledgments}\label{sec:ak} This work has been partially supported by the Wallenberg Autonomous Systems and Software Program (WASP) funded by Knut and Alice Wallenberg Foundation and by the TrustFull project financed by the Swedish Foundation for Strategic Research.\looseness=-1

\bibliographystyle{ieeetr}
\IEEEtriggeratref{15}
\bibliography{bibliography/ref}
\end{document}